\documentclass[letterpaper]{article}

\usepackage{amsmath}
\usepackage{mathtools}
\usepackage{braket}
\usepackage{amssymb}

\usepackage[T1]{fontenc}

\usepackage{geometry}
\usepackage{setspace}

\usepackage[backend=biber,url=false,doi=false]{biblatex}

\usepackage{hyperref}
\usepackage{cleveref}

\usepackage{graphicx}
\usepackage{float}
\newfloat{scheme}{htbp}{los}
\floatname{scheme}{Scheme}
\floatname{chart}{Chart}
\newfloat{graph}{htbp}{loh}

\usepackage{chemformula} 
\usepackage[version = 4]{mhchem} 

\usepackage{authblk}

\title{Programmable photonic waveguide arrays: opportunities and challenges}

\author[1,*]{Yang Yang}
\author[1,2]{Akram Youssry}
\author[1,3,$\dagger$]{Alberto Peruzzo}
\affil[1]{Quantum Photonics Laboratory and Centre for Quantum Computation and Communication Technology, RMIT University, Melbourne, VIC 3000, Australia}
\affil[2]{School of Electrical Engineering and Telecommunications, UNSW Sydney, Sydney, NSW 2052, Australia}
\affil[3]{Quandela, Massy, 91300, France}
\affil[*]{yangyang00770@gmail.com}
\affil[$\dagger$]{alberto.peruzzo@gmail.com}

\begin{document}

\maketitle

\begin{abstract}
The rising complexity of photonic applications, ranging from quantum computing to neuromorphic processing, has driven the demand for highly programmable and scalable photonic integrated circuits. While mesh-based architectures built from Mach-Zehnder interferometers have enabled significant advances, their reliance on beam splitting and light bending introduces optical loss, fabrication challenges, and scalability bottlenecks.  Continuously lateral-coupled integrated waveguide arrays (WAs), by contrast, offer compact systems with no direct free-space analogs, but their static nature has limited their utility. Recently, programmable waveguide arrays (PWAs) have emerged as a promising alternative, combining the Hamiltonian richness of WAs with tunable control. This perspective outlines the conceptual foundations, recent progress, and future potential of PWAs across quantum simulation, photonic neural networks, topological photonics, and nonlinear optics. We examine the theoretical and practical challenges of modeling, fabrication, and control, and propose PWAs as a next-generation architecture for compact, reconfigurable photonic processors.
\end{abstract}

\section*{Keywords}
programmable photonic circuits, waveguide arrays, universal multiport interferometers, photonic matrix multiplication, Hamiltonian engineering, integrated quantum photonics


\section{Introduction}

The rapid progress in photonic technologies over the past few decades has transformed both research and industry, driven by advances in materials (e.g., silicon photonics~\supercite{chen2018emergence,rahim2018open,blumenthal2018silicon,quack2023integrated}, III-V semiconductors~\supercite{smit2012moore,smit2019past,klamkin2018indium}), fabrication techniques (e.g., direct laser writing~\supercite{deubel2004direct,szameit2010discrete}, thin-film technologies~\supercite{maszara1988bonding,colinge2012silicon,zhu2021integrated}), and applications ranging from quantum computing~\supercite{wang2020integrated} to topological photonics~\supercite{RevModPhys.91.015006}, quantum cryptography~\supercite{pirandola2020advances}, and artificial intelligence~\supercite{wetzstein2020inference}.

\begin{figure}
    \centering
    \includegraphics[width=0.42\columnwidth]{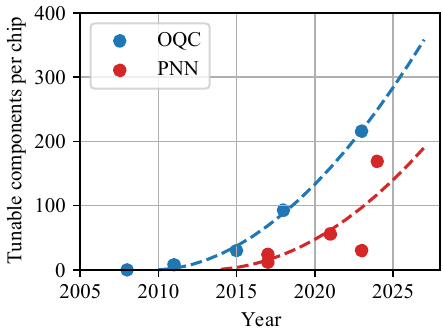}
    \caption{\textbf{Increasing complexity of programmable photonic circuits for advanced applications, quantified by the number of tunable phase shifters per chip.} The development trend is indicated by an exponential function fitting represented by dashed lines. Although recent milestones have been achieved primarily with MZI mesh networks, the continued exponential growth in complexity highlights the demand for alternative architectures. PWAs may provide a more scalable and compact solution to continue this trend. Notable milestones in the field of optical quantum computing (OQC) include the first static CNOT gate (a static device with 0 tunable components in 2008)~\supercite{politi2008silica}, the first programmable quantum processor (8 tunable components in 2011)~\supercite{shadbolt2012generating}, the first universal linear circuits (30 tunable components in 2015)~\supercite{carolan2015universal}, the first large-scale circuit (93 tunable components in 2018)~\supercite{wang2018multidimensional}, and a recent large-scale circuit (216 tunable components in 2023)~\supercite{bao2023very}. Key demonstrations in the field of photonic neural networks (PNN) include the first demonstration of unscrambling light (12 tunable components in 2017)~\supercite{annoni2017unscrambling}, the first demonstration of deep learning (24 tunable components in 2017)~\supercite{Shen_2017}, the first implementation of complex-valued neural networks (56 tunable components in 2021)~\supercite{zhang2021optical}, the first realization of backpropagation training for deep learning (30 tunable components in 2023)~\supercite{doi:10.1126/science.ade8450}, the first demonstration of a fully-integrated coherent optical neural network (169 tunable components in 2024)~\supercite{bandyopadhyay2024single}.}
    \label{fig:num_components}
\end{figure}

Among these advances, programmable photonic circuits have become a key enabler of flexible and scalable devices~\supercite{Bogaerts2020}, with Mach-Zehnder interferometer (MZI) based mesh structures~\supercite{Reck_1994, Clements_2016, Perez_2017} being widely adopted for applications in microwave photonics~\supercite{Marpaung_2019}, neuromorphic computing~\supercite{shastri2021photonics}, and quantum information processing~\supercite{carolan2015universal}. These mesh architectures rely on discrete 2×2 beam splitters and phase shifters, adapted from free-space optics, and have scaled rapidly in complexity over the last two decades, as illustrated in Figure.~\ref{fig:num_components}, which summarizes the growth in circuit size and component count across different photonic platforms. However, they are fundamentally limited by the need to bend light paths, contributing to optical loss and limiting circuit density.

In contrast, photonic waveguide arrays (WAs) offer a unique platform that implements ``always-on'' Hamiltonians through evanescent coupling in a fixed spatial geometry. These structures have no direct analogs in free-space optics and have enabled pioneering studies in light transport~\supercite{PhysRevLett.81.3383,PhysRevE.66.046602,christodoulides_discretizing_2003,fleischer2003observation}, Bloch oscillations~\supercite{morandotti1999experimental,PhysRevLett.83.4752}, Anderson localization~\supercite{schwartz2007transport,lahini2008anderson}, quasicrystal dynamics~\supercite{PhysRevLett.110.076403,levi2011disorder}, and topological photonics~\supercite{RevModPhys.91.015006,longhi2009quantum}. Their use has also expanded to classical and quantum information processing, including quantum walks~\supercite{peruzzo2010quantum}, matrix multiplication~\supercite{zhou2022photonic}, and quantum state generation~\supercite{doi:10.1126/sciadv.aat3187,chapman2023quantum,doi:10.1126/science.aau4296}. Traditional waveguide arrays have long enabled studies in light dynamics and quantum simulations. However, their static nature, with parameters fixed at fabrication, imposes severe limitations on scalability, adaptability, and reusability across applications.

\begin{figure}
    \centering
    \includegraphics[width=0.5\columnwidth]{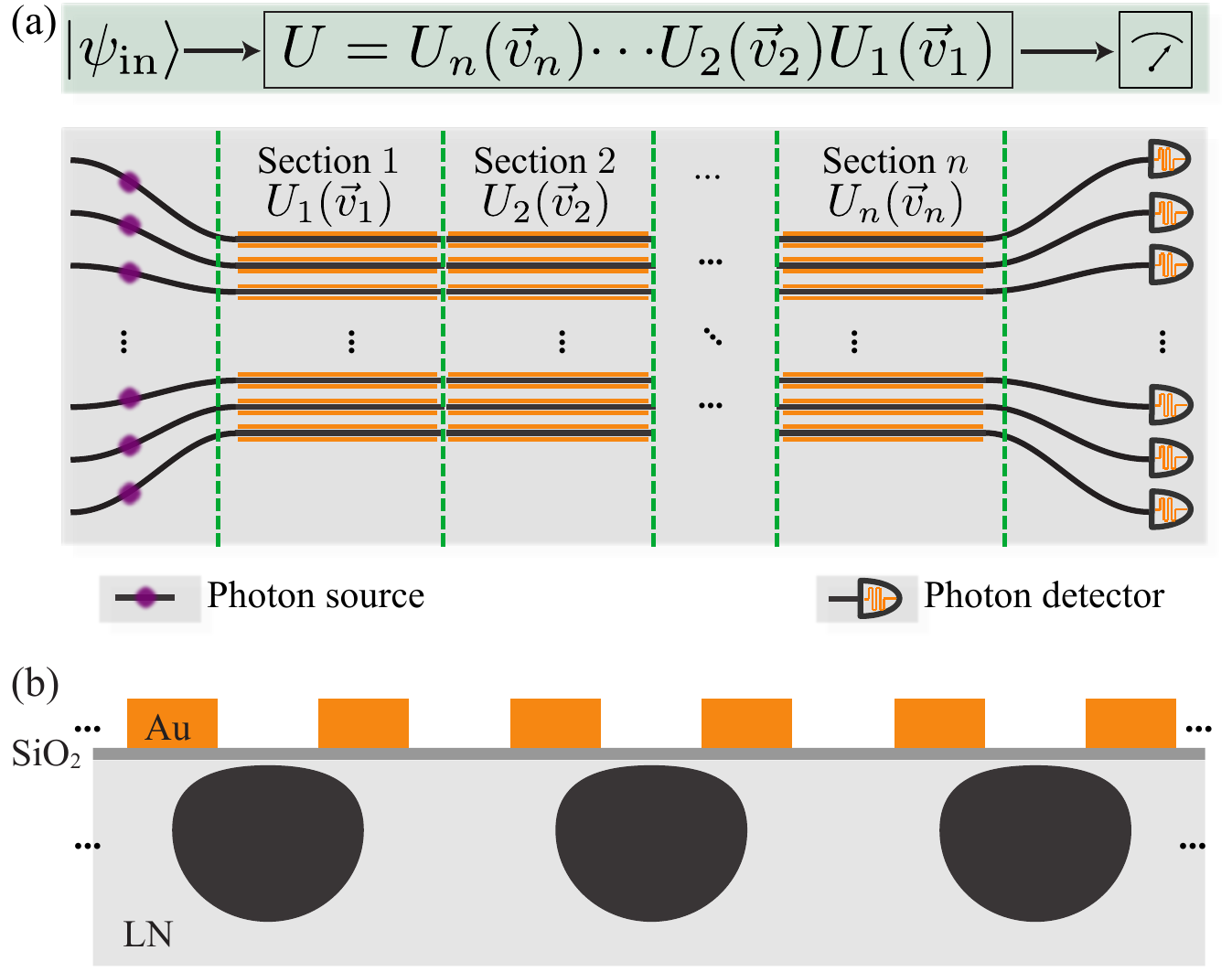}
    \caption{\textbf{Schematics of an integrated PWA architecture}. (a) The architecture incorporates photon sources and detectors with potential for integration as well as cascaded $n$ multisection PWAs. The black lines represent optical waveguides, and the orange lines indicate control connections. Each section of PWA has independent tuners and implements a $N$-dimensional unitary transformation matrix $U_k(\vec{v}_k)$, where $\vec{v}_k$ is the control parameter applied to the tuners in the $k$-th section. The entire architecture implements an overall unitary transformation matrix $U$. For a given input state $\ket{\psi_\text{in}}$, the output state is calculated as $\ket{\psi_\text{out}}=U\ket{\psi_\text{in}}$. (Note: This architecture can be used for both classical and quantum light.) (b) Cross-section of an example PWA featuring three waveguides (black areas indicate the optical modes) based on the device recently demonstrated in Ref~\supercite{Yang_2024,yang2024programmablesingle}. The device is fabricated using annealed proton-exchange waveguides on x-cut lithium niobate~\supercite{lenzini2015anisotropic}. Gold microelectrodes functioning as microtuners were patterned on top of the silicon dioxide buffer layer above the waveguides. Electric fields are generated by applying control pulses to the electrodes and are confined by the shielding effect of neighboring electrodes, effectively eliminating crosstalk~\supercite{10011218}. This ensures precise control of individual Hamiltonian terms and their corresponding unitary transformations.}
    \label{fig:vision_v4}
\end{figure}

We propose a new paradigm: programmable waveguide arrays (PWAs)~\supercite{Yang_2024}, which merge the continuous coupling of static arrays with the reconfigurability of mesh networks~\supercite{yang2024programmablesingle}. This hybrid architecture opens unprecedented avenues in quantum computing, neuromorphic photonics, and topological protection~\supercite{Price_2022,10621808}.
Figure~\ref{fig:vision_v4} illustrates the conceptual architecture of programmable waveguide arrays, highlighting how local control elements enable dynamic reconfiguration of coupling and on-site parameters. Beyond active reconfiguration, PWAs also leverage structural programmability: the coupling between adjacent guides can be preset during fabrication by controlling their separation, enabling fine-grained tuning of the device's Hamiltonian even before dynamic control is applied.
Bridging concept and implementation, recently developed PWAs introduce tunability through these local controls~\supercite{Yang_2024}, allowing direct programming of  Hamiltonians and the realization of arbitrary unitary transformations without resorting to light-bending geometries. This shift marks a departure from conventional MZI-based processors, offering a compact and intrinsically analog alternative that could scale more favourably for large or dynamically evolving systems. Furthermore, PWA-based architectures have recently been shown to be universal~\supercite{Youssry2024controllability}, capable of emulating both time-independent and time-dependent Hamiltonians with reduced optical loss and minimal footprint.

In this perspective, we highlight the emerging potential of continuously-coupled integrated programmable waveguide arrays as a versatile interdisciplinary platform. We start with an introduction to their operating principles and examine the associated challenges in modeling and design (Section.~\nameref{section:PWA}). This is followed by a discussion on device characterization and control (Section.~\nameref{section:characterzation_control}). We then survey application domains where PWAs could unlock new capabilities (Section.~\nameref{section:applications}). Finally, we conclude with a future outlook, presenting a roadmap and identifying open research directions (Section.~\nameref{section:conclusion}).

\begin{figure}
    \centering
    \includegraphics[width=1.0\columnwidth]{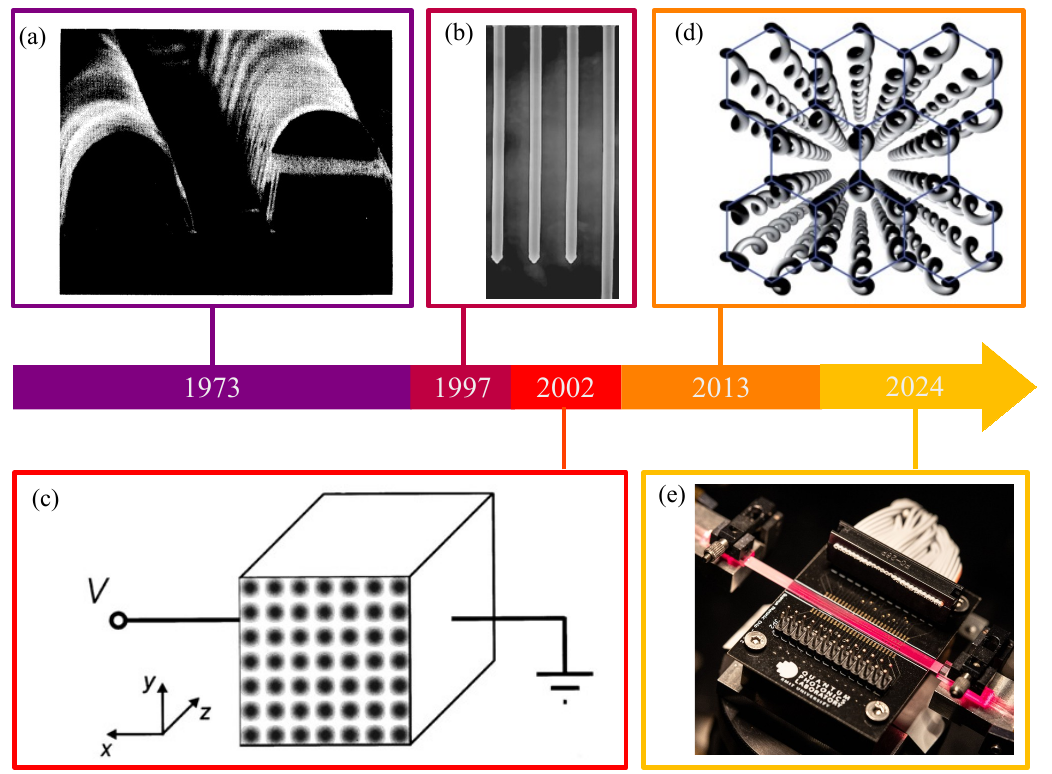}
    \caption{\textbf{Evolution of waveguide array systems.} (a) The first experimental demonstration of a directional coupler fabricated in GaAs in 1973. Reproduced with permission from ref.~\supercite{10.1063/1.1654468}, Copyright 1973 AIP Publishing.     (b) The first demonstration of a nonlinear waveguide array fabricated on an AlGaAs platform in 1997. Reproduced with permission from ref.~\supercite{millar1997nonlinear}, Copyright 1993 Optica Publishing Group. (c) The two-dimensional optically induced waveguide array in a bulk photorefractive crystal in 2002. Reproduced with permission from ref.~\supercite{PhysRevE.66.046602}, Copyright 2002 American Physical Society. (d) The two-dimensional helical waveguides fabricated using femtosecond laser writing technology in 2013. Reproduced with permission from ref.~\supercite{rechtsman_2013}, Copyright 2013 Nature Publishing Group UK London. (e) The fully electro-optically programmable waveguide array fabricated in bulk lithium niobate used in recent demonstrations~\supercite{youssry2024experimental,Yang_2024,yang2024programmablesingle} (Red light shines through the chip to highlight the device). Reproduced with permission from James Thomas, Oneill Photographic. Copyright 2024 James Thomas. RMIT logo reproduced with permission from RMIT University. Copyright 2026 RMIT University.} 
    \label{fig:evolution_WA}
\end{figure}

\section{Device modeling and design}\label{section:PWA}

In this section, we introduce the fundamentals of PWA and examine the challenges in PWA modeling and design, along with potential strategies to address them. We begin with a brief overview of waveguide array systems, followed by two mathematical modeling approaches for PWA devices. Lastly, we review photonic material platforms suitable for realizing PWAs and address essential considerations for circuit layout and waveguide geometry design.

\subsection{Overview of continuously coupled waveguide systems}

Traditional waveguide arrays are composed of evanescently coupled waveguides integrated on photonic platforms. These devices enable optical waves to discretize and propagate through coupled channels, a concept first proposed to achieve light discretization~\supercite{lederer2008discrete}. Since the initial theoretical proposal in 1965~\supercite{Jones:65} and the first experimental demonstrations in 1973~\supercite{10.1063/1.1654468}, WAs have been realized on various material platforms, with steady advances in device size and circuit complexity. The smallest PWA architecture is a reconfigurable directional coupler--two coupled waveguides augmented by a local control element that enables dynamic tuning of their coupling. The field is now advancing toward more complex configurations involving multi-waveguide and multi-section PWAs.
The conceptual evolution of waveguide arrays is illustrated in Figure.~\ref{fig:evolution_WA}, tracing the progression from rigid, fabrication-limited lattices to dynamically tunable architectures that define programmable waveguide arrays. This transition marks a shift from passive photonic design to active, software-defined photonic functionality.

Light propagation in WAs can be described by Hamiltonian time evolution, wherein the physical and geometric properties of the device define the individual terms of the Hamiltonian and evolution time. The input-output transformation implemented by the respective device can be observed by measuring the output state given a specific light input state or by obtaining fluorescence microscopy images through implanted color centers~\supercite{szameit2007quasi}. WAs are typically static, application-tailored, and single-purpose devices.  

Programmable waveguide arrays are assembled by adding precisely positioned microtuners to the static WA structures. These microtuners placed around the waveguides allow modification of the device's Hamiltonian by applying control pulses. The design of microtuners can vary depending on the selected tuning mechanism and material platform, which is discussed in the following content. An ideal PWA enables precise control over each term in its Hamiltonian, thereby rendering it fully programmable. Consequently, the unitary transformation implemented by such devices can be controlled. A conceptual schematic of a PWA, adapted from the recently demonstrated device~\supercite{Yang_2024}, is shown in Figure.~\ref{fig:vision_v4}.

\subsection{Theoretical model} \label{section:basic}

A cascaded PWA architecture with $n$ sections implements a unitary transformation matrix:
\begin{align}
    U = \prod_{k=1}^{n}U_k(\vec{v}_k),
\end{align}
where $\vec{v}_k$ is the set of control voltages applied on the $k-$th section, $U_k$ is the transformation implemented by each section, and the product is defined in reverse order:
\begin{align}
    \prod_{k=1}^{n}U_k(\vec{v}_k) =  U_n(\vec{v}_k) \cdots U_3(\vec{v}_3) U_2(\vec{v}_2) U_1(\vec{v}_1).
\end{align}
Each section has a propagation length $L_k$, and can be described using a Hamiltonian $H_k$, such that
\begin{align}
    U_k(\vec{v}_k)=e^{-iH_k(\vec{v}_k)L}
    \label{1.1.2}.
\end{align}
Each of those Hamiltonians can be expressed as a tri-diagonal matrix using discrete differential equations~\supercite{christodoulides_discretizing_2003,lederer2008discrete}:
{\footnotesize
\setlength{\arraycolsep}{1pt} 
\medmuskip = 0.1mu 
\begin{equation}
H_k(\vec{v}_k) = \begin{bmatrix} 
    \beta_{1}(\vec{v}_k) & C_{1,2}(\vec{v}_k) & 0 &  \dots & 0 & 0\\
    C_{2,1}(\vec{v}_k) & \beta_{2}(\vec{v}_k) & C_{2,3}(\vec{v}_k) &  \dots & 0 & 0\\
    0 & C_{3,2}(\vec{v}_k) & \beta_{3}(\vec{v}_k) & \dots & 0 & 0\\
    \vdots & \vdots & \vdots & \ddots & \vdots &\vdots\\
    0 & 0 & 0  & \dots & \beta_{N-1}(\vec{v}_k) & C_{N-1,N}(\vec{v}_k) \\
    0 & 0 & 0 & \dots & C_{N,N-1}(\vec{v}_k) & \beta_{N}(\vec{v}_k) \\
    \end{bmatrix}
    \label{1.1.1}
\end{equation}}
where $C_{a,b}$ is the coupling coefficient between waveguides $a$ and $b$, $\beta_i$ is the propagation constant of waveguide $i$ and $N$ is the number of waveguides. When designing a static device Hamiltonian, the propagation constants are typically set by adjusting the refractive index of the waveguides~\supercite{itoh2006ultrafast}, whereas the coupling coefficients are determined by the waveguide refractive index and the distance between them~\supercite{yariv1991optical}.

\color{black}

This modeling method is a widely used approach that assumes a tridiagonal Hamiltonian with real-valued terms. The parameters are obtained generally based on a two-waveguide system. Although this method is subject to several assumptions, it remains a justified approximation for the static design of devices and serves as a starting point.

Note that achieving a coupling coefficient of $C_{a,b} = 0$ is not possible with any known optical material or arbitrary control mechanism, such as voltage tuning. The only feasible way to achieve $C_{a,b} = 0$ is by physically separating the waveguides, as is done in MZI-based technologies. Furthermore, setting $\beta_i = 0$ is not valid, as it would eliminate the mode and prevent light from being guided, and $\beta_i < 0$ is not considered, as the model assumes forward-only photon propagation~\supercite{lifante2003integrated}. The constraint $C_{a,b}\neq0$ adds complexity to the study of controllability and was only recently addressed~\supercite{Youssry2024controllability}. Notably, this constraint is absent in solid-state systems, which simplifies the controllability study of those platforms.

An alternative method is to model a device based on optical supermodes solved using Finite-Difference Time-Domain simulations as shown in~\supercite{yang2024fixed}. 
This also applies to asymmetric modes in waveguide arrays and does not rely on the assumption of a tri-diagonal Hamiltonian. It offers a more accurate approximation for static device design, especially in a more compact WA system where waveguides are placed closer together, making higher-order coupling non-negligible. 

While the dependence of the Hamiltonian on the tunable control parameters can be modeled theoretically or through multiphysics simulations, the exact relationship in a real device may vary due to fabrication uncertainties. Therefore, a proper device characterization procedure is essential to accurately determine this dependence, as discussed later. 

The universality of the architecture is achieved based on the following theorem ~\supercite{Youssry2024controllability}:\\
\textit{Let $\tilde{K}=\frac{1}{6}\left(2d^3 - 3 d^2 + d\right)$, and $N\in \mathbb{Z}^{+}$. For $d>2$, an arbitrary $U \in \text{SU}(d)$ matrix can be decomposed into $K=8\tilde{K} N$ cascaded sections, each described by a tridiagonal Hamiltonian with strictly-positive elements. The error between the original matrix $U$ and the decomposition $V$ scales asymptotically as $\|U-V\| = O\left(\frac{\tilde{K}}{N}\right)$. For $d=2$, the decomposition is exact with $K=4$ sections at most.}

A rigorous proof of this result can be found in~\supercite{Youssry2024controllability}. The theorem guarantees that with enough sections, the error between the target and actual unitaries can be made arbitrarily small. However, the bounds are not tight, and thus, in practice, we can achieve high-fidelity gates with a reasonable number of sections, as will be demonstrated later. Therefore, there are three key requirements for achieving universality:
\begin{enumerate}
    \item A sufficient number of cascaded and independently controllable PWA sections;
    \item Tunability of on-site propagation constants and inter-waveguide coupling coefficients;
    \item A varied physical length for each PWA section.
\end{enumerate}

\color{black}

While PWAs can be described at the Hamiltonian level, their practical scalability is strongly constrained by material platforms, waveguide geometry, and circuit layout. The dense integration of tunable elements amplifies the impact of propagation loss and tuner-induced crosstalk, motivating the following discussions in Section.~\nameref{section:material} and~\nameref{section:reconfigurable_modeling_design}.

\subsection{Material platforms}\label{section:material}


Although a wide range of material platforms is available for integrated photonics, it remains unclear which are best suited for realizing scalable and high-performance programmable waveguide arrays. Note that Figure.~\ref{fig:vision_v4} (b) is an illustrative example of a possible PWA implementation, rather than a prescription for a specific material platform or tuning mechanism. In this subsection, we discuss key challenges associated with waveguide-array fabrication and tuning. We also examine candidate material platforms in terms of the design trade-offs they present for PWA implementations, to motivate future investigations in this emerging area.

Low-propagation-loss waveguides are crucial for achieving reconfigurable WAs because photon loss significantly affects many applications. The smoothness of the waveguide surface plays a significant role in minimizing the propagation loss. Various fabrication approaches have been used for photonic WAs, with direct laser writing of waveguides~\supercite{davis1996writing,szameit2007control,szameit2010discrete} and lithographic and etching-based technologies being the most widely adopted methods~\supercite{chen2018emergence,rahim2018open,munoz2019foundry}. Direct laser writing enables the fabrication of two-dimensional WAs~\supercite{meany2015laser} and helical waveguides~\supercite{rechtsman_2013}. Lithographic and etching-based methods, on the other hand, provide low-loss waveguides~\supercite{blumenthal2018silicon} and support large-scale integration~\supercite{bao2023very,shekhar2024roadmapping}. However, they are often constrained to planar waveguide arrays, limiting their ability to implement the time evolution of one-dimensional quantum walks only.

To minimize control crosstalk, a crucial requirement for achieving PWAs is the ability to tune device parameters both locally and efficiently. Some WA devices have demonstrated limited reconfigurability, primarily relying on the thermo-optic effect~\supercite{PhysRevLett.83.4752,hoch2022reconfigurable}. This effect is commonly used in photonic platforms such as silica~\supercite{meany2015laser}, silicon~\supercite{fathpour2015emerging,chen2018emergence,rahim2018open}, and silicon nitride~\supercite{munoz2019foundry}. However, the diffusive heat propagation in materials makes the thermo-optic effect non-ideal for tuning device refractive index locally~\supercite{10011218,https://doi.org/10.1002/lpor.202000024}. The plasma dispersion effect represents an alternative tuning scheme~\supercite{reed2010silicon}, which is employed in indium phosphide photonics~\supercite{smit2019past,klamkin2018indium,shih2012indium,doerr2015silicon}. However, this results in higher optical losses as the tuning efficiency increases~\supercite{1073206}, making it unsuitable for applications that are sensitive to optical losses~\supercite{bartolucci2023fusion} or large-scale circuits.   
An alternative tuning mechanism can utilize micro-electro-mechanical systems (MEMS) technology in silicon photonics, which has recently attracted significant attention. Its capability to displace waveguides at the nanoscale in various configurations makes it well-suited for precisely controlling the coupling strength between waveguides. However, MEMS systems are sensitive to mechanical vibrations, which reduce the fidelity of the operation. Additionally, to achieve a phase shifter in MEMS systems, a second coupled waveguide must work as a loading to create an effective mode index change, which introduces unavoidable light loss~\supercite{8847362,quack2023integrated}. PWAs based on the electro-optic effect have recently been demonstrated in bulk lithium niobate~\supercite{Yang_2024}. Transitioning to thin-film lithium niobate can yield a reduced footprint, low propagation loss~\supercite{siew2018ultra,krasnokutska2018ultra}, and enhanced electro-optic efficiency~\supercite{zhu2021integrated}. This includes better electrode placement, high-speed operation, and a low driving voltage~\supercite{Zhang:21, Krasnokutska:2019_Scientific_Reports}. The integration of on-chip light sources~\supercite{Jankowski:20,White:19,10.1063/1.5054865} and detectors~\supercite{Lomonte:2021} offers the potential to enhance scalability, reduce coupling losses, and eliminate the need for bending areas in fan-in and fan-out regions. However, materials such as aluminum nitride~\supercite{dietrich2016gaas} and gallium arsenide~\supercite{lu2018aluminum,10045080}, which can leverage electro-optic effect, suffer from lower electro-optic coefficients and higher propagation losses.  
Barium titanate (BTO), the optical material with the highest known electro-optic coefficient, could also be used to fabricate PWA. Although BTO waveguides can be fabricated, the technology is still in its early stages, leading to high optical losses~\supercite{petraru2002ferroelectric,czornomaz2022bto}. Alternatively, other approaches like BTO-on-SiN offer promising solutions, but they demand high-quality thin films and carefully engineered interfaces between BTO and other materials~\supercite{alexander2024manufacturable}. Other promising directions for investigation include exploring phase-change and magneto-optic materials~\supercite{wuttig2017phase,shoji2018waveguide,radford2025inverse}, leveraging laser-induced photorefractive effects~\supercite{sturman2021photovoltaic}, and developing innovative hybrid approaches~\supercite{elshaari2020hybrid,moody20222022,pelucchi2022potential}.

\subsection{Circuit layout and waveguide geometry optimization} \label{section:reconfigurable_modeling_design}
A PWA device implements a controlled version of the static Hamiltonian, providing independent control of the parameters in the Hamiltonian. The mathematical and physical requirements must be considered when designing PWA devices. Ideally, a single section of a PWA should be capable of implementing arbitrary unitary transformations by tuning its Hamiltonian parameters using its tuners. However, owing to the physical constraints imposed on the control signal (such as limited amplitude to avoid physical damage to the device) and finite device length, achieving this with a single section can be challenging. Consequently, it is crucial to account for the available unitary space accurately under given physical and control constraints. This mathematical problem, known as controllability, is particularly challenging owing to the nonlinear relationship between the Hamiltonian parameters and corresponding unitary transformation matrices.  

The design of the waveguide array geometry and choice of material platforms are crucial for determining the feasible parameter space that a device can explore. The waveguide array geometry determines the static Hamiltonian and influences the design of the tuner. On the other hand, the choice of the material platform determines the tuning mechanism and impacts the tuner design. The most challenging aspect of tuner design is the precise placement of tuners relative to waveguides, as they often need to be positioned between waveguides, where the distance between coupled waveguides is typically less than a few micrometers. Moreover, tuners may need to be positioned at a distance to maintain the high index contrast of the waveguide and to prevent control signal crosstalk, especially when the tuners are made from photon-absorptive materials. Fabrication and design constraints, including wafer size, which may limit the device length and the total number of waveguides, as well as the geometry of the tuner nanostructure, which can impact the control signal transmission, must be carefully considered. Additionally, the effective coupling length of a PWA, which determines the Hamiltonian evolution time, must be carefully designed by considering the static coupling strengths of adjacent waveguides, tunability of the tuners, and target applications. All these design considerations combined will largely define the potential unitary space that a PWA device can achieve.    

\begin{figure}
    \centering
    \includegraphics[width=0.5\columnwidth]{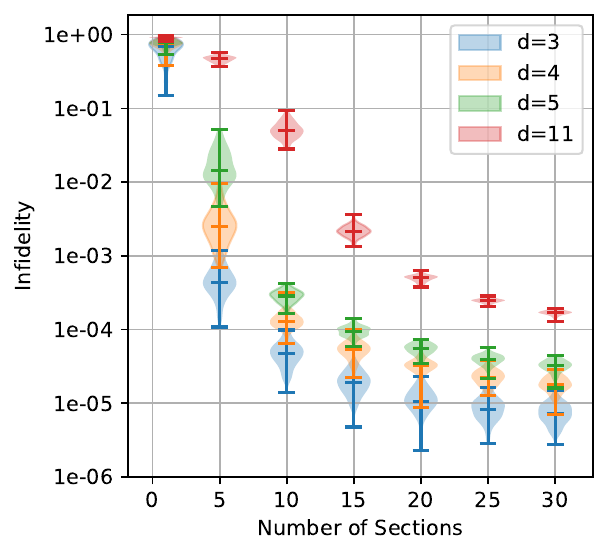}
    \caption{\textbf{Infidelity of 100 numerically optimized Haar-random unitaries as a function of the number of PWA sections.} The infidelity decreases with increasing section count. $d$ denotes the number of modes in the programmable waveguide array. Reprinted with permission under a Creative Commons Attribution 4.0 License from ref.~\supercite{Youssry2024controllability}. Copyright 2025 IOP Publishing.}
    \label{fig:unitary_coverage}
\end{figure}

In practice, the reachable unitary space can be expanded by cascading multiple independently programmable PWA sections. Although the tunability of each section is limited by finite interaction length and bounded control amplitudes, sequential application of controlled Hamiltonians allows increasingly expressive transformations through their accumulated evolution. This trend is quantitatively illustrated in Figure.~\ref{fig:unitary_coverage}, which shows that increasing the number of sections systematically reduces the infidelity for Haar-random target unitaries across different dimensionalities. These results highlight a clear engineering trade-off between device footprint, tuning complexity, and achievable unitary coverage.

\section{Characterization and control}\label{section:characterzation_control}

Device characterization is essential for a wide range of applications~\supercite{lahini_quantum_2018,chapman2023quantum,yang2024programmablesingle}. Moreover, achieving controllable PWAs requires a controller capable of generating precise control parameters for specific unitary transformations based on the results of device characterization.

\subsection{Device characterization}\label{section:characterzation}

The process of device characterization involves measuring the device to establish a mapping between its control parameters and the corresponding Hamiltonian or unitary transformation matrix~\supercite{RevModPhys.91.045002}. Characterizing a PWA device to explore its achievable unitary space involves modeling the device and fitting the model using numerous measurements. These measurements can be performed using either classical light~\supercite{rahimi2013direct,youssry2020modeling,hoch2023characterization} or quantum light via HOM measurements~\supercite{peruzzo2011multimode,laing2012super,dhand2016accurate}, similar to the process used in mesh scheme characterizations. However, the mathematical approach to constructing the device model differs, as PWAs do not allow completely decoupled waveguides. Therefore, machine learning-based methods are commonly employed~\supercite{youssry2020modeling,youssry2024experimental}, typically relying on a predefined physical model, such as the one discussed in Section.~\nameref{section:basic}. Other approaches, such as optimization techniques~\supercite{tillmann2016unitary,spagnolo2017learning} or supervised learning algorithms~\supercite{kuzmin2021architecture}, can also be adapted to characterize PWAs.

However, the cost of data-driven methods can be substantial, as the complexity of characterizing a PWA device grows with the increasing number of waveguides and tuners in the system. As the number of tuners increases, so does the number of measurements required for characterization, making scaling an open question. Consequently, approximation methods may be necessary. Furthermore, owing to the nonlinear dependence of the time evolution unit on the tuning parameters, the choice of sampling methods can be critical, potentially affecting the required number of measurements. Because this task requires a number of measurements that grow with the dimensions, classical light is more feasible than quantum light characterization because measurements with quantum light require a significant amount of time to accumulate photon statistics, and laser light characterization completely predicts the quantum process of linear photonic devices. Additionally, other concerns in device control, such as pulse shaping aimed at stabilizing the device state, must be considered.  

\begin{figure}
    \centering
    \includegraphics[width=1.0\columnwidth]{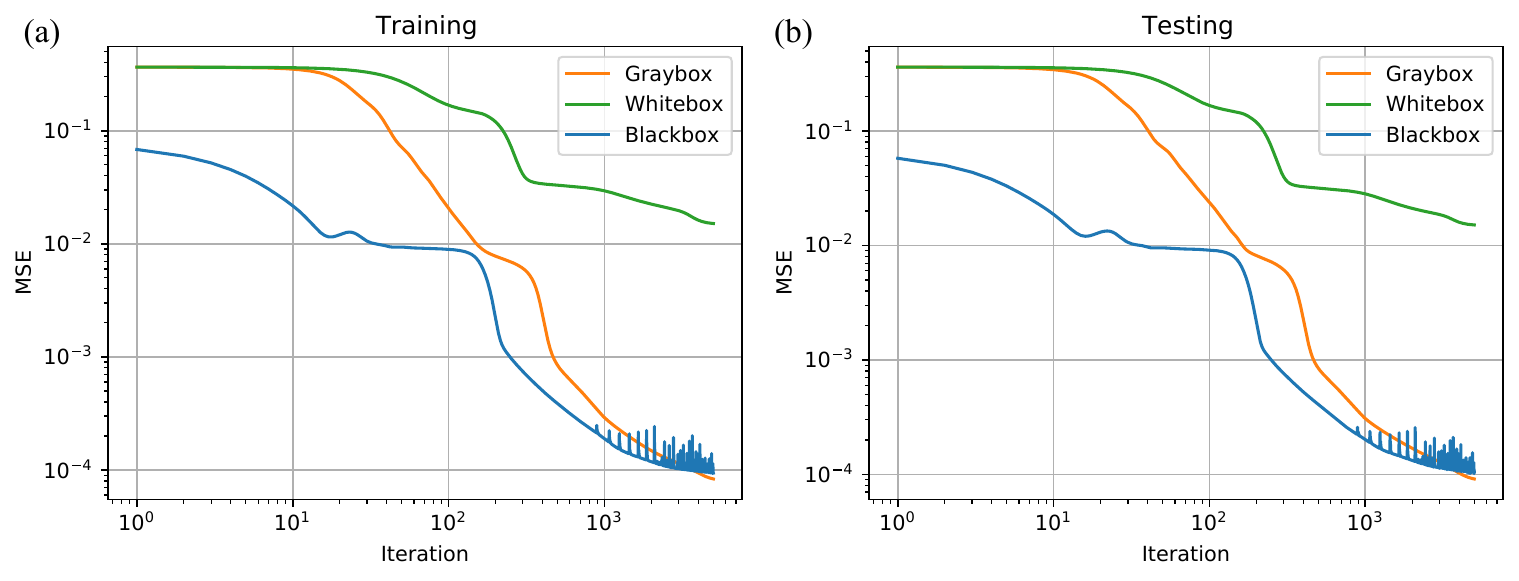}
    \caption{\textbf{Experimental training and testing convergence of data-driven machine learning models.} The experiment is performed on a three-mode lithium-niobate-based programmable waveguide array. The dataset consists of 7000 samples for training and 1000 samples for testing. Each sample contains the measured output power distribution for a given input mode under randomly chosen voltages applied to the control electrodes. Reprinted with permission under a Creative Commons Attribution 4.0 International License from ref.~\supercite{youssry2024experimental}. Copyright 2024 Nature Publishing Group UK.    
    }
    \label{fig:ml_convergence}
\end{figure}

Figure.~\ref{fig:ml_convergence} illustrates representative convergence behavior for different machine learning models applied to experimental data acquired for a three-mode PWA chip, where the mean-squared error (MSE) decreases systematically with iteration number during both training and testing. The results highlight that incorporating partial physical structure into the model (e.g., a physics-informed or a “graybox” approach) can significantly improve convergence speed and final accuracy compared to purely black-box methods, while outperforming purely whitebox models that are usually subject to approximations and/or strong assumptions. These trends demonstrate that data-driven calibration and control of PWAs is practically feasible, while also emphasizing the trade-offs between model fidelity, optimization complexity, and robustness that must be considered in scalable implementations. Details of this experiment can be found in~\supercite{youssry2024experimental}.

\subsection{Device control}

Once an effective model is established, the next step is to design a controller that can precisely manage the model to achieve the desired target functionalities. This requires inverse mapping of the tuning parameters to their corresponding unitary operations. Studies have shown that this can be achieved using machine learning methods~\supercite{youssry2020modeling,youssry2024experimental, bohlin2006practical}. A significant challenge in this process is that the experimental resources needed to accurately train a machine learning model may become impractical.

A study with a 3-PWA system~\supercite{youssry2024experimental} revealed that the method in Section.~\nameref{section:basic} is incomplete for modeling a fabricated physical device. This work shows that by relaxing the tri-diagonal Hamiltonian assumption and allowing for a complex-valued Hamiltonian, the fitting performance of the model is improved significantly. This is due to the fact that the fabricated device is non-homogeneous along the propagation direction due to either material non-homogeneity or imperfect electrode alignment. This suggests that a fabricated device deviates notably from the model of an ideal device. 

A linear relationship between the device’s control parameters and the tridiagonal Hamiltonian model presented in~\supercite{Yang_2024} has been shown to be valid to a certain extent. Moreover, a model-free approach has also been demonstrated~\supercite{yang2024programmablesingle}. This approach involves building lookup maps based on the measurements obtained from different control configurations. Reinforcement learning methods have also been demonstrated to solve this control problem~\supercite{fouad2024modelfreedistortioncancelingcontrol}. Furthermore, algorithms that enable progressive self-configuration using transparent detectors in MZI networks have been extensively applied~\supercite{miller2013self_oe,miller2013establishing,miller2013reconfigurable,miller2013self_invited}. Such algorithms have been extended to the Clements' scheme~\supercite{PhysRevApplied.18.024019} and a more compact version recently~\supercite{hamerly2022asymptotically}. Developing similar algorithms to characterize PWA networks could be an intriguing area for further research.

\section{Translating potential applications to the programmable paradigm}\label{section:applications}


The following sections examine key domains where PWAs could significantly expand functionality, efficiency, or usability, even if full demonstrations are yet to be realized. Each application area is introduced through its static WA heritage or MZI-based implementation, followed by an exploration of how programmability could enable new capabilities or resolve existing bottlenecks. While the applications discussed here do not introduce new physical effects, they highlight the potential of programmable waveguide arrays as a unified and reconfigurable platform for accessing, consolidating, and systematically exploring a broad class of Hamiltonian-based photonic phenomena previously demonstrated using static devices. This provides a common conceptual framework for researchers across different applications and expertise domains.

\begin{figure}
    \centering
    \includegraphics[width=1.0\columnwidth]{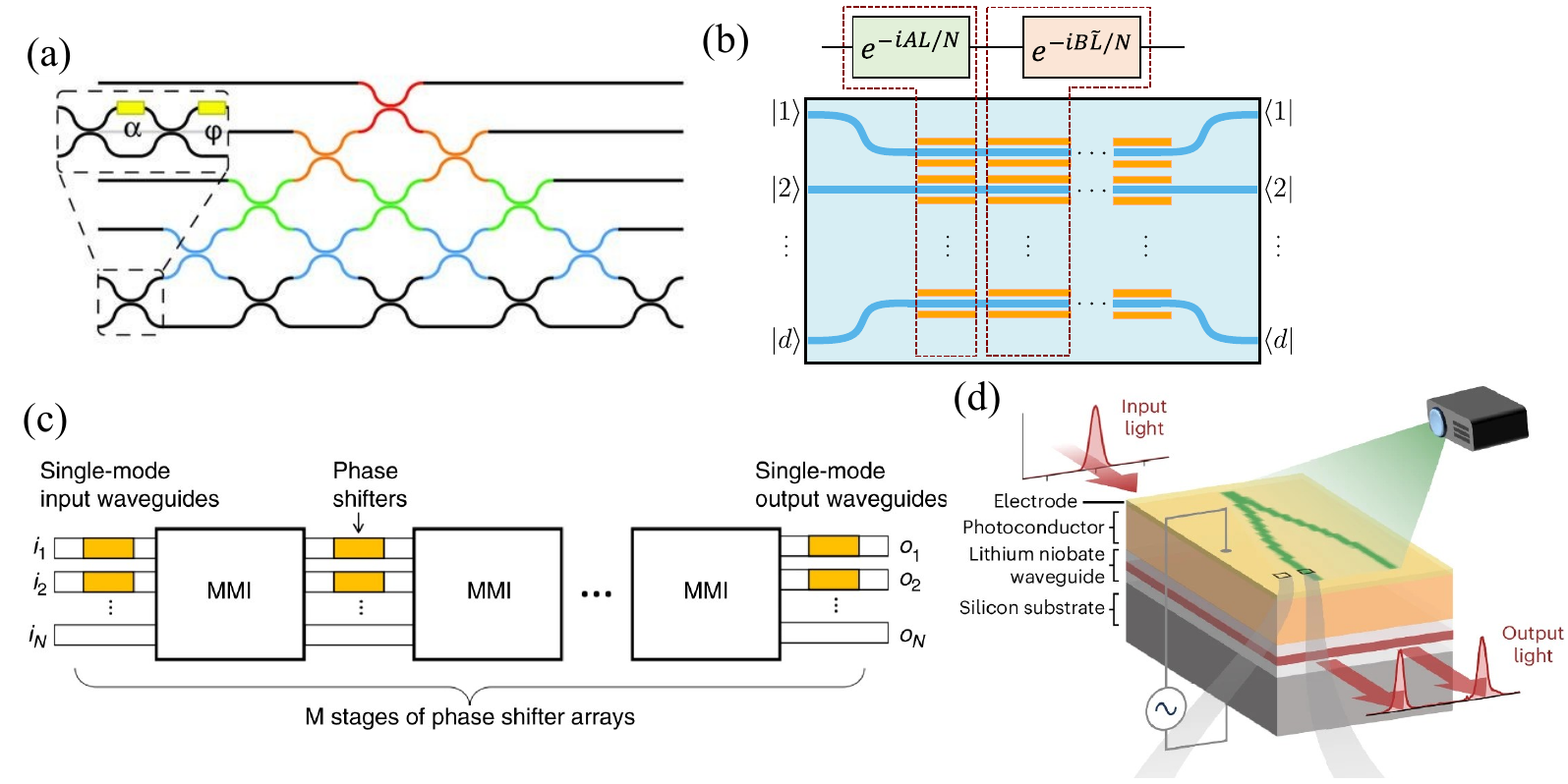}
    \caption{\textbf{Programmable photonic processor architectures.} (a) Mach–Zehnder interferometer (MZI)–based universal multiport interferometer in a triangular configuration. Reproduced with permission from ref.~\supercite{carolan2015universal}, Copyright 2015 American Association for the Advancement of Science. (b) Top-view schematic of a PWA architecture. Reprinted with permission under a Creative Commons Attribution 4.0 License from ref.~\supercite{Youssry2024controllability}. Copyright 2025 IOP Publishing. (c) Architecture based on cascaded multi-mode interferometers (MMIs) and phase shifters. Reproduced with permission from ref.~\supercite{tang2017integrated}, Copyright 2012 IEEE. (d) Programmable multimode wave propagation in a multimode waveguide or scattering region. Reprinted with permission under a Creative Commons Attribution 4.0 International License from ref.~\supercite{onodera2025arbitrary}. Copyright 2025 Nature Publishing Group UK.}
    \label{fig:architectures}
\end{figure}

\begin{figure}
    \centering
    \includegraphics[width=1.0\columnwidth]{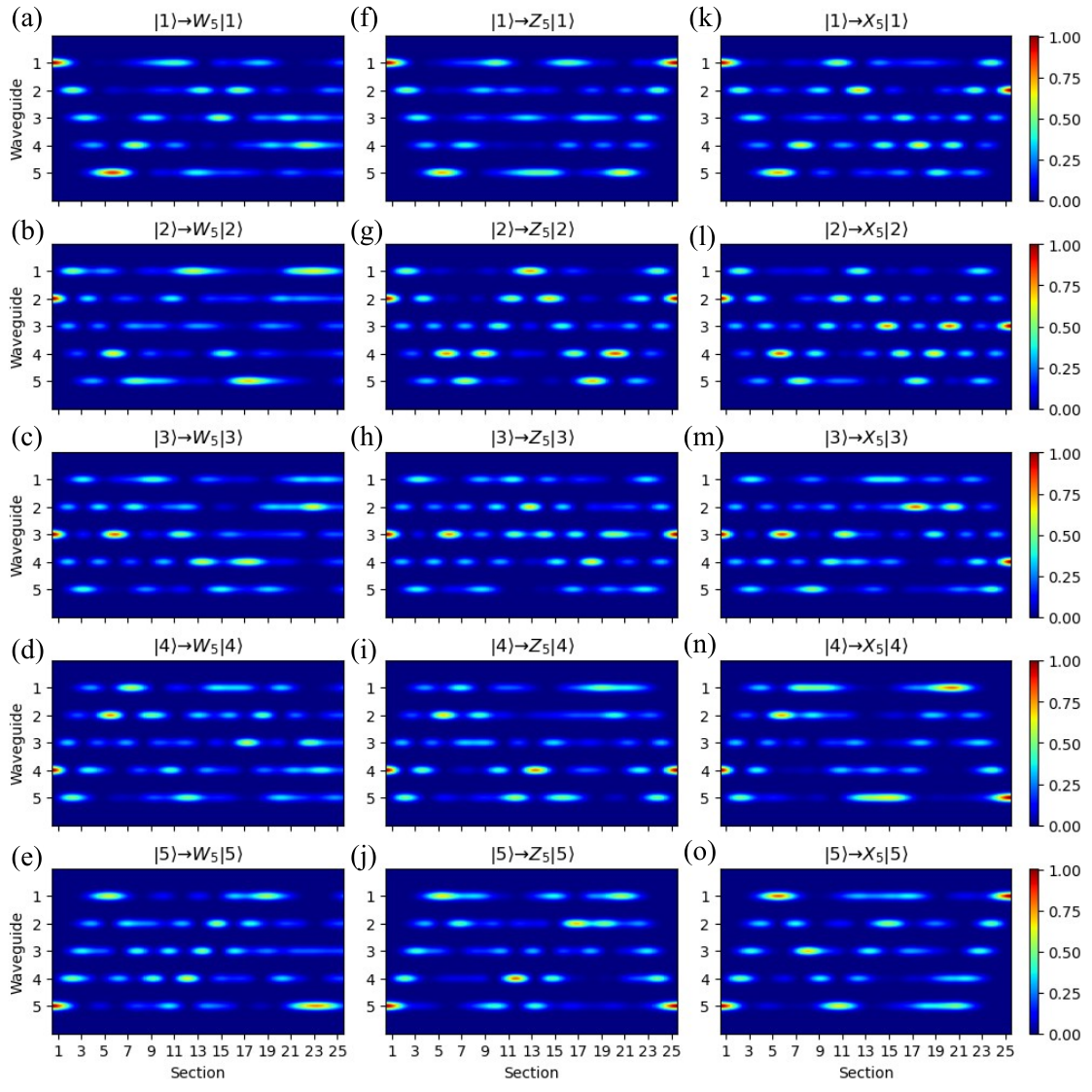}
    \caption{\textbf{Optical field evolution simulation for a 25-section, 5-mode PWA.} The simulated applications include (a) - (e) discrete Fourier transform (DFT) gate $W_5$, (f) - (j) the clock gate $Z_5$, and (k) - (o) the shift matrix $X_5$. Reprinted with permission under a Creative Commons Attribution 4.0 License from ref.~\supercite{Youssry2024controllability}. Copyright 2025 IOP Publishing.}
    \label{fig:optical_field_evolution}
\end{figure}

\subsection{Universal multiport interferometers}\label{Universal multiport interferometers}

\paragraph*{Application Overview:} Universal multiport interferometers enable arbitrary linear unitary matrix transformations between multiple optical channels. They have extensive applications in both classical and quantum information processing~\supercite{Bogaerts2020,zhou2022photonic}, which will be discussed in more detail in the following content. In addition to MZI-based mesh structures (Figure.~\ref{fig:architectures} (a)), which are considered discrete or digitized, there has been growing interest in leveraging continuously coupled waveguides or multimode interferometers (MMIs) as error-robust and more scalable alternatives to realize universal multiport interferometers~\supercite{petrovic_new_2021,kim2025programmable}.
In 2017, a hybrid structure based on cascading MMIs and phase shifters was proposed to implement unitary transformations as illustrated in Figure.~\ref{fig:architectures} (c)~\supercite{tang2017integrated}. This structure was further investigated in~\supercite{saygin_robust_2020}, where several transformations were numerically studied. Experimental demonstrations of such an architecture have also been performed~\supercite{tang2018reconfigurable,tang2021ten}. Enhanced scalability and error tolerance have also been demonstrated for large-scale circuits~\supercite{tanomura_scalable_2022}. The realization of multiport transformations using structures with partially programmable waveguide arrays has also been proposed~\supercite{skryabin_waveguide-lattice-based_2021}. 
Moreover, programmable multimode interferometers have recently emerged as a platform for implementing arbitrary unitary transformations by exploiting global mode mixing within multimode waveguides or scattering regions, as illustrated in Figure.~\ref{fig:architectures} (d)~\supercite{larocque2021universal,wu2023lithography}. These architectures relax constraints on waveguide spacing and layout, thereby simplifying fabrication and enabling high modal density. While such platforms are well-suited for applications such as photonic neural networks~\supercite{onodera2025arbitrary}, their control is typically indirect and optimization-based. As a result, they do not rely on an explicit unitary decomposition and lack a direct mapping to tight-binding Hamiltonians with spatially localized and independently addressable coupling terms.


\paragraph*{Potential of PWAs:}
Structures based on cascaded PWAs take full advantage of continuously coupled waveguides in integrated photonics (Figure.~\ref{fig:architectures} (b)). They have also been recently rigorously proven to be universal~\supercite{Youssry2024controllability}, addressing a long-standing question raised in previous studies. The work introduced a decomposition algorithm for realizing any unitary matrix using PWAs with guaranteed error bounds. For $U(2)$, a maximum of 4 sections are needed, which is the exact decomposition. The required number of sections $n$ scales as $O(N^3 k)$, with the error scaling as $O(N^3/k)$, where $k$ is the Trotterization number. While this result suggests a high number of sections may be required to achieve low fidelities, the study demonstrates that, in practice, the number of sections needed is significantly smaller than the theoretical bound. By leveraging realistic materials parameters and numerical optimization to compile the unitary, these requirements become practically achievable. 
Figure.~\ref{fig:optical_field_evolution} shows a simulation of the evolution of the optical field inside a 5-mode waveguide device, programmed to implement (a) a discrete Fourier transform (DFT) gate $W_5$, (b) a clock gate $Z_5$, (c) the shift gate $X_5$~\supercite{Youssry2024controllability}. In this simulation, 25 sections are used, and the voltages are obtained via numerical optimization. The simulation parameters are given in~\supercite{Youssry2024controllability}.
Therefore, cascaded PWAs are functionally equivalent to the widely adopted MZI-mesh structures~\supercite{Reck_1994, Clements_2016}, while they eliminate the need for bending. This unique feature can potentially lead to significant reductions in optical losses and circuit footprint.

\subsection{Quantum computing}

\paragraph*{Application Overview:}
Quantum computers promise to solve computational problems more efficiently than classical systems~\supercite{ladd2010quantum}. In 1982, Feynman proposed using controllable quantum systems to simulate the dynamics of other systems~\supercite{feynman2018simulating}, now known as analog quantum computing~\supercite{buluta2009quantum,johnson_2014}. This approach allows the simulator’s Hamiltonian to be controlled to match the target system, offering greater error robustness, less demanding hardware, and better scalability compared to universal digital architectures~\supercite{doi:10.1080/23746149.2018.1457981,RevModPhys.86.153,trivedi2024quantum}. Waveguide arrays, which naturally implement time evolution governed by their Hamiltonian, have been used to simulate optical analogs of solid-state systems~\supercite{PhysRevLett.109.106402}. By varying waveguide coupling along the propagation length and integrating control electrodes to tune coupling strength, PWAs can simulate time-dependent Hamiltonians for analog quantum simulations. 
Integrated photonics is a promising platform for gate-based quantum circuits~\supercite{slussarenko2019photonic,moody20222022}. Digital quantum computing~\supercite{deutsch1989quantum}, where qubits evolve through single- and two-qubit gates, is universal~\supercite{lloyd1996universal} and can be implemented in photonic systems using mesh structures of $2 \times 2$ beam splitters and phase shifters~\supercite{Reck_1994, Clements_2016, Perez_2017}. While static, continuously coupled waveguides have been used to implement quantum gates~\supercite{lahini_quantum_2018, chapman2023quantum}, digital quantum circuits require reconfigurable systems. 

\paragraph*{Potential of PWAs:}
Recently, controllable single-qubit gates have been demonstrated~\supercite{yang2024programmablesingle}, and cascaded PWAs offer a pathway to universal digital quantum circuits~\supercite{Youssry2024controllability}.
A PWA-based digital quantum simulator could implement Trotterization steps by discretizing time evolution into small unitary operations~\supercite{lloyd1996universal}. By tuning the waveguide coupling and interaction parameters, this setup can approximate the continuous evolution of a system’s Hamiltonian, simulating digital quantum gates effectively. This approach provides a scalable route to implement digital quantum algorithms using cascaded waveguide arrays.
By supporting both analog and digital quantum information processing, cascaded PWAs offer a unique capability in photonics. Digital-analog quantum computing (DAQC), which combines the strengths of both approaches, has been proposed to address the challenges of the noisy intermediate-scale quantum era~\supercite{PhysRevA.101.022305}. Using analog methods for continuous evolution and digital precision for error correction, DAQC enhances efficiency, error tolerance, and scalability, offering a promising path to solve complex quantum problems with fewer resources.\

\subsection{Photonic neural networks}

\paragraph*{Application Overview:} Photonic neural networks (PNNs) are emerging as a transformative platform for accelerating machine learning, leveraging the unique properties of light to achieve ultra-high-speed, low-latency, and energy-efficient computations~\supercite{shastri2021photonics,liao2023integrated,peserico2023integrated}. At their core, PNNs rely on photonic matrix multiplication to perform matrix-vector products, a fundamental operation in neural network processing~\supercite{zhou2022photonic}. By leveraging quantum light as input, PNNs can enable quantum machine learning~\supercite{biamonte2017quantum,Bogaerts2020} and other quantum information processing tasks such as Hamiltonian simulation, quantum information compression, and one-way quantum repeaters~\supercite{steinbrecher2019quantum}. PNNs have been successfully demonstrated across various photonic platforms~\supercite{Shen_2017, Harris:18,annoni2017unscrambling,roques2020heuristic,flamini2020photonic,zhang2021optical,doi:10.1126/science.ade8450}, with MZI meshes being among the most widely used architectures. Other approaches, such as cascaded MMI couplers and phase shifter arrays, also provide viable alternatives. 

\paragraph*{Potentials of PWAs:} Programmable waveguide arrays present a compelling alternative hardware platform for implementing photonic neural networks. Compared to MZI-based or hybrid MMI–phase shifter architectures, PWAs may more easily scale to larger input dimensions, support multi-wavelength operations, and reduce overall optical losses. Furthermore, due to the universality and robustness of PWAs, the training of PWA-based PNNs can be streamlined with a simplified calibration process. Additionally, while PWAs are fundamentally linear photonic platforms, their continuously coupled waveguide geometry can facilitate the integration of nonlinear activation elements for neural networks through optical nonlinear materials or detector co-integration~\supercite{bandyopadhyay2024single}.


\subsection{Topological photonics}

\paragraph*{Application Overview:} 
Topological photonics has become a vibrant and rapidly evolving field over the past decade, drawing inspiration from the discovery of quantum Hall effects and topological insulators in condensed matter physics~\supercite{RevModPhys.91.015006, Price_2022, PhysRevA.108.040101}. WAs, with their ability to emulate tight-binding Hamiltonians, have proven to be powerful simulators of electron behavior in such systems~\supercite{Gao_Xu_Yang_Zwiller_Elshaari_2024}. This has enabled the experimental exploration of a wide range of topological phenomena, including the first demonstration of parity-time (PT) symmetry in optics~\supercite{ruter2010observation} and photonic Floquet topological insulators~\supercite{rechtsman_2013}. These topological systems usually exhibit intrinsic robustness to disorder~\supercite{khanikaev2024topological}, making them promising for practical applications in areas such as optical isolation~\supercite{el2015optical, PhysRevApplied.11.034045}, communications~\supercite{manzalini2020topological}, spintronics~\supercite{he2019topological}, and even quantum computing. A wide range of topologically protected quantum states has been realized in WA-based platforms~\supercite{rechtsman2016topological, gorlach2017topological, doi:10.1126/science.aau4296, doi:10.1126/sciadv.aat3187, blanco2019topological, wang2019topologically, wang2019topological_2, wang2019quantum, wang2019direct, hashemi2025topological}, affirming the potential of these systems for fundamental science and future technologies.

\paragraph*{Potentials of PWAs:}
PWAs are poised to significantly expand the capabilities of topological photonic systems. Unlike static WAs, PWAs allow the exploration of multiple topological phases within a single device. This programmability also opens the door to dynamically probing robustness by introducing controlled disorder or tunable defects~\supercite{Yang_2024}. Additionally, with reconfigurable coupling and phase shifts, PWAs could provide a flexible testbed for implementing topological optical isolators, as well as exploring emergent phenomena in disordered or time-varying systems~\supercite{titum2015disorder, stutzer2018photonic}.

Furthermore, the modularity of PWAs suggests a potential route toward cascading multiple two-dimensional arrays with integrated dynamic modulation. Such architectures may serve as compact alternatives to helically curved designs, enabling photonic simulations of time-dependent Hamiltonians while mitigating bending-induced losses~\supercite{doi:10.1126/science.aaz8727, mukherjee2020observation}. With these features, PWAs could be well-positioned to serve as a flexible and scalable platform for next-generation topological photonic devices.

\subsection{Nonlinear phenomena}

\paragraph*{Application Overview:} 
Nonlinear optical effects have significantly expanded the functional landscape of integrated photonics, particularly in coupled waveguide systems. Early work began with nonlinear directional couplers~\supercite{jensen1982nonlinear} and soon extended into WAs~\supercite{Christodoulides:88}, where nonlinear interactions gave rise to spatial solitons~\supercite{millar1997nonlinear,PhysRevLett.81.3383}, nonlinear Bloch oscillations~\supercite{morandotti1999experimental,PhysRevLett.83.4752}, and complex dynamics in two-dimensional lattices~\supercite{PhysRevE.66.046602,fleischer2003observation}. These studies have contributed to advances in pulse shaping~\supercite{hudson2008nonlinear}, all-optical switching~\supercite{fleischer2005spatial}

In recent years, the integration of nonlinear optics with topological photonics has opened even more compelling avenues. Nonlinear WAs have enabled the realization of self-localized topological edge states~\supercite{lumer2013self,leykam2016edge}, topological insulator lasers that maintain robustness under fabrication disorder~\supercite{zhou2017optical}, and chip-scale generation of topologically protected entangled photon pairs via SPDC and four-wave mixing~\supercite{solntsev2012spontaneous,PhysRevX.4.031007,solntsev2017path,smirnova2020nonlinear}. These breakthroughs highlight the growing potential of nonlinear photonics as a foundation for both classical and quantum optical technologies.

\paragraph*{Potentials of PWAs:}
PWAs offer a promising path to dynamically enhance and control nonlinear phenomena in integrated photonics. Unlike static WAs, PWAs enable real-time control over coupling and phase parameters, making it possible to dynamically shape nonlinear interactions. This tunability provides a new level of flexibility for controlling soliton dynamics, optimizing nonlinear Bloch oscillations, and adaptively shaping pulses for ultrafast optics or optical computing. Moreover, when built from or integrated with materials exhibiting strong second- or third-order nonlinearities, PWAs could become tunable platforms for efficient wavelength conversion, light filtering, and dynamic routing in nonlinear circuits.

The ability to combine nonlinearity with topological protection in a programmable platform is particularly powerful. With PWAs, one could study the emergence and control of nonlinear topological edge states under varying interaction strengths, or explore regimes where topological protection competes or cooperates with nonlinear localization. Reconfigurability also enables new approaches to simulate nonlinear-driven phase transitions and dynamically explore higher-order topological phenomena that are otherwise difficult to access with static systems.

\subsection{Other potential emerging applications}
Additional potential applications of precisely engineered device Hamiltonians and lengths include quantum walks~\supercite{PhysRevLett.100.170506,peruzzo2010quantum,grafe2016integrated}, simulating non-Markovian giant atom decay~\supercite{longhi2020photonic} and higher-dimensional systems~\supercite{maczewsky_synthesizing_2020}, emulating and harnessing non-Abelian braiding ~\supercite{iadecola2016non,zhang2022non}, generating W states~\supercite{PhysRevA.87.013842,grafe2014chip}, and achieving perfect state transfer~\supercite{chapman2016experimental}. The reconfigurability of PWAs offers a unified platform for realizing these applications with enhanced precision, scalability, and adaptability. Time-dependent Hamiltonians could potentially be realized by cascading PWAs, with each section implementing a different Hamiltonian~\supercite{rechtsman_2013,mukherjee2020observation,doi:10.1126/science.aaz8727}. Moreover, the incorporation of gain and loss in PWAs can be achieved by applying various control mechanisms to waveguide tuners, including the creation of defect waveguides~\supercite{trompeter2003tailoring,jorg2017dynamic}. This setting enables the realization and control of non-Hermitian dynamics~\supercite{PhysRevLett.100.103904}, unlocking phenomena such as parity-time symmetry~\supercite{ruter2010observation} and topological phase transitions~\supercite{zeuner2015observation}, as well as applications like the generation of topologically protected bound states~\supercite{weimann2017topologically}. 

\begin{figure}
    \centering
    \includegraphics[width=0.4\columnwidth]{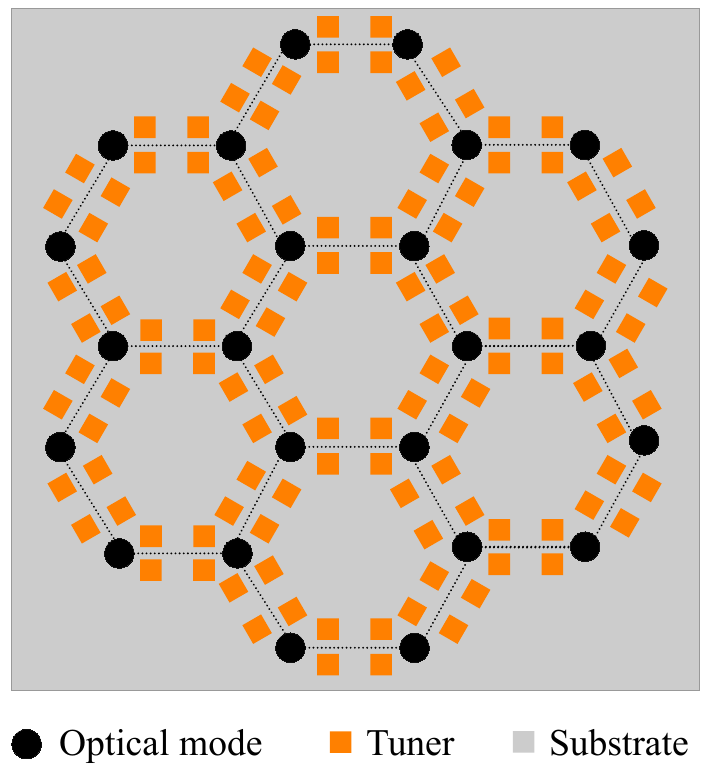}
    \caption{\textbf{Cross-section of a two-dimensional PWA concept design.} The optical modes are arranged in a hexagonal pattern~\supercite{rechtsman_2013}, as indicated by dashed lines. Four tuners between neighboring optical modes control their coupling coefficients, while six tuners placed around an optical mode adjust their propagation coefficients. The substrate is made of an ideal isotropic material, such as one with a homogeneous high electro-optic coefficient in all directions.}
    \label{fig:3d_pwa}
\end{figure}

\section{Conclusion and outlook}\label{section:conclusion}

Programmable waveguide arrays combine the Hamiltonian richness and compact form factor of static waveguide arrays with the versatility of reconfigurable photonic platforms. Their ability to implement “always-on” Hamiltonians with tunable parameters enables a wide range of applications, including quantum walks, analog simulations, topological photonics, and photonic neural networks. Unlike mesh-based programmable architectures, PWAs eliminate the need for bending and beam splitting, offering potential advantages in optical loss, circuit footprint, and fabrication tolerance.

Recent demonstrations have shown that cascaded PWAs are capable of implementing arbitrary unitary transformations, including time-dependent Hamiltonian evolution, providing a flexible framework for quantum information processing, reconfigurable signal processing, and hybrid digital-analog computation. This versatility may unlock high-dimensional encoding schemes in communications~\supercite{cozzolino2019high}, robust sensing with programmable dynamics~\supercite{degen2017quantum,taha2024photonics}, and new protocols for quantum and neuromorphic computing.

The theoretical framework presented in~\supercite{Youssry2024controllability} establishes the universality of PWAs under a constrained tridiagonal Hamiltonian model (Eq.~\ref{1.1.1}), though practical performance is inherently limited by fabrication variability and first-order approximations. Incorporating higher-order coupling terms~\supercite{yariv1991optical} or transitioning to multi-dimensional geometries may reduce the number of sections required to achieve target fidelities. These considerations naturally motivate the exploration of two-dimensional programmable waveguide arrays, where increased connectivity can reduce architectural overhead while expanding the accessible Hamiltonian space.

Two-dimensional PWA structures could be enabled by integrating liquid metals as tunable elements~\supercite{daeneke2018liquid}, as illustrated in Figure.~\ref{fig:3d_pwa}. Compared to one-dimensional PWAs, two-dimensional PWA structures provide a device Hamiltonian that can be more flexibly and precisely engineered through additional tuners distributed in the available spatial degrees of freedom. This could enable control over couplings even beyond nearest neighbors and along multiple spatial directions, facilitated by waveguide bending or multi-planar layouts. Such richer connectivity allows the implementation of more complex Hamiltonians, including higher-order interactions and effective long-range couplings, which are difficult or inefficient to realize in strictly one-dimensional architectures. As a result, certain unitary transformations or lattice models may be implemented with fewer cascaded sections in a two-dimensional PWA than in an equivalent one-dimensional configuration. Moreover, two-dimensional PWAs naturally enable the emulation of higher-dimensional physical systems and topological phases, providing direct access to phenomena that cannot be faithfully captured in one-dimensional lattices~\supercite{maczewsky_synthesizing_2020}. While the practical realization of two-dimensional PWAs introduces additional challenges in fabrication complexity, tuning density, and crosstalk management, their enhanced connectivity and increased expressive power make them a promising pathway toward scalable and functionally dense programmable photonic processors.

To advance towards scalable and deployable PWA-based systems, the field must now establish clear quantitative performance benchmarks. We identify four immediate priorities: 
\begin{enumerate}
    \item Increasing the number of precisely programmable waveguides and sections beyond current prototypes, typically limited to 3–11 waveguides and a single section, remains a major design and fabrication challenge. 
    \item Defining the achievable level of control per waveguide and per section, including independent tuning of coupling and phase parameters, is essential for establishing the functional limits of PWA architectures.
    \item Systematically reporting key performance indicators--such as insertion loss, control voltage range, and coupling tunability--is needed to enable meaningful cross-platform comparisons (e.g., with MZI-based meshes).
    \item Evaluating the engineering gap between academic demonstrations and deployable systems across target applications will help identify the dominant bottlenecks and guide technology maturation strategies.
\end{enumerate}

These benchmarks will be essential for advancing the technology, comparing material platforms, and determining whether PWAs are ready to support real-world applications or remain limited to proof-of-concept experiments.

Future work may also explore the implementation of Haar-random unitaries in PWAs, which form the basis of complexity conjectures in photonic quantum computing~\supercite{aaronson2011computational,kruse2019detailed}. As the number of tunable parameters grows, the need for scalable characterization and control techniques becomes increasingly urgent. Advances in data-driven modeling, in-situ feedback, and high-density tuning elements will be instrumental in overcoming these challenges.

Overall, programmable waveguide arrays offer a promising new paradigm for reconfigurable integrated photonics. As control fidelity, circuit complexity, and integration improve, PWAs are poised to play a central role in scalable quantum technologies, adaptive optics, and photonic information processing.

\section*{Acknowledgements}

The authors thank Tim Weiss and Hamed Arianfard for valuable discussions and feedback on the manuscript. AP acknowledges an RMIT University Vice-Chancellor’s Senior Research Fellowship and a Google Faculty Research Award. This work was supported by the Australian Government through the Australian Research Council under the Centre of Excellence scheme (No: CE170100012).

\section*{Competing interests}
The authors declare no competing interests.

\section*{Associated Content}
Yang Y, Youssry A, Peruzzo A. Programmable photonic waveguide arrays: opportunities and challenges. 2025, arXiv:2502.12385, https://doi.org/10.48550/arXiv.2502.12385 (accessed May 28, 2026).

\printbibliography

\end{document}